# Refugee Law Lab Working Paper

**Title:** Gender, Sexual Orientation, and the One-Year Refugee Bar: A Computational Study of Bill C-12 and Canadian Refugee Adjudication

**Author:** Sean Rehaag

**Abstract:** In March 2026, Bill C-12 amended the *Immigration and Refugee Protection Act* to make refugee claims ineligible for referral to the Refugee Protection Division where they are made more than one year after the claimant's first entry to Canada. Affected claimants are diverted to a Pre-Removal Risk Assessment (PRRA), which is typically decided by an immigration officer on a paper record, with no guaranteed oral hearing, no appeal on the merits, and no automatic stay of removal pending judicial review. The government defended the bar partly on the premise that genuine refugees claim promptly. Critics countered that late claims are concentrated among vulnerable groups, particularly people fleeing persecution based on gender, sexual orientation, gender identity, or gender expression (SOGIE).

This article tests those competing premises. Using artificial intelligence tools, checked through human review, it examines nearly 29,000 published decisions of the Refugee Protection Division, the Refugee Appeal Division, and the Federal Court from 2006 to 2025. The results of this study demonstrate that, first, while most claimants claim within a year, roughly one in nine in the published record does not, and gender and SOGIE claims are significantly overrepresented among these late claims. Second, among late claims, refusals of gender and SOGIE claims are more likely than refusals of other claims to be overturned on appeal or judicial review. Late claims on other grounds, by contrast, fare somewhat worse on review than timely claims. The one-year bar therefore removes the hearing, the appeal, and the stay from the late claimants for whom those safeguards have been doing the most work. The article recommends repealing the bar or, failing that, adopting exemptions, procedural protections, and greater transparency in PRRA decision-making.

# Gender, Sexual Orientation, and the One-Year Refugee Bar: A Computational Study of Bill C-12 and Canadian Refugee Adjudication

Sean Rehaag[1]

## 1. Introduction

A young man from Nigeria comes to Canada on a study permit to pursue a Bachelor of Engineering degree. He spent his adolescence being careful about how he looked at other boys, careful about who got close. In Nigeria, where same-sex sexual intimacy is criminalized, he never said the word "gay" about himself to anyone, not even to himself. In Canada, slowly, that changes. He joins a queer student group at school. He starts dating a man. During his final year of studies, a cousin in Toronto sees him walking into a gay bar with a group of men and tells his family back home. The threats begin almost immediately. His mother warns him that it is not safe for him to come home and that his father, a police officer, wants him dead. Recent immigration law changes mean he will no longer qualify for permanent residence after graduating, so he asks his school's legal clinic how he can stay in Canada. He is surprised to learn that it is possible to ask for refugee protection based on persecution involving sexual orientation. He asks how refugee claims work.

The answer he receives depends on timing. Until recently, his claim would have been decided through Canada's full refugee determination system, including an oral hearing before an independent tribunal, an appeal on the merits, and protection from removal while the process ran its course. Under amendments enacted through Bill C-12, however, claims made more than one year after a claimant's first entry to Canada are no longer eligible for that system. What he gets instead is a Pre-Removal Risk Assessment (PRRA), which is an evaluation conducted by an immigration officer, typically based on a paper review, with far fewer procedural safeguards.

Public debates about this one-year bar turn on competing empirical premises. The government maintains that genuine refugees claim promptly and that a firm deadline will therefore mostly screen out people misusing the asylum system. Refugee advocates counter that some well-founded claims take longer than a year to surface and that late claims concentrate among specific groups, especially people fleeing persecution based on gender and sexual orientation.

---

[1] Associate Professor, Osgoode Hall Law School, York University; Director, Refugee Law Lab; Co-Director, Access to Algorithmic Justice; Project Director, Algorithmic Justice for Refugees Research Network. The author is grateful to Simon Wallace, Clifford McCarten and Arghavan Gerami for helpful comments on prior drafts and to Youdon Tenzin for able research assistance. The article draws on research supported by the Social Sciences and Humanities Research Council of Canada, the Law Foundation of Ontario, the Canada Foundation for Innovation, and the Ontario Research Fund. Generative AI tools were used in preparing the article, including for background research, data extraction, coding, drafting and editing. All citations in the article were manually checked.

These are empirical claims that can be tested, and that is what this article sets out to do.

The article uses computational methods to extract information from thousands of Refugee Protection Division (RPD), Refugee Appeal Division (RAD) and Federal Court decisions, including whether the claim was made more than a year after the claimant's first entry to Canada, and the grounds on which the claim was based. Statistical analysis is then undertaken on the resulting dataset to answer two questions. First, are claims made more than a year after the claimant's first entry to Canada more likely than timely claims to involve allegations of gender-based persecution or of persecution based on sexual orientation, gender identity, or gender expression (SOGIE)? Second, among those late claims, are refusals of gender and SOGIE claims more likely than refusals of other claims to be overturned on appeal or judicial review? If the answer to both questions is yes, then advocates for refugees appear to be right in saying that the one-year bar channels many claims by particularly vulnerable claimants that are especially in need of error correction into a refugee determination stream with reduced error correction potential. Along the way, the study also puts the government's own premise to the test, by measuring how common late claims are in the published record and how they fare at each level of the refugee determination system.

The article proceeds as follows. Part 2 sets out the context for the research, including the pressures that produced the 2026 reforms, the legislative roadmap, the mechanics of the one-year bar, the government's justifications, the critiques, the American experience with a similar rule, and the absence of Canadian evidence on the competing premises. Part 3 sets out the methodology and limitations of the study, as well as its findings. Part 4 explores the implications of the findings and offers recommendations for reform, and Part 5 concludes.

## 2. Context

### *2.1 The road to the 2026 reforms*

Recent years have seen growing preoccupation with the number of refugee claims made in Canada.

While Canada hosts only a tiny fraction of the total number of refugees around the world,[2] and while many countries with far fewer resources host far more refugees per capita,[3] the number of claims made in Canada has increased recently. By 2013, after years of the former Conservative government's policies and rhetoric aimed to deflect and discourage asylum seeking,[4] the number of annual claims dropped to a low of approximately 10,000 claims, from numbers that fluctuated between 20,000 and 35,000 per year in the 10 years prior. The number of claims rose steadily after that. By 2019, the annual figure had climbed to roughly 58,000. After a brief pandemic-era dip, claims further increased to approximately 94,000 in 2022, 147,000 in 2023, and 174,000 in 2024.[5]

---

[2] UNHCR, *Global Trends: Forced Displacement in 2025* (Geneva: UNHCR, 2026), online: <https://www.unhcr.org/media/global-trends-2025-report> at 5-6 (reporting 117.8 million forcibly displaced people worldwide at the end of 2025).
[3] Ibid at 3 (reporting that low- and middle-income countries host 68 percent of the world's refugees and that a single country, Colombia, hosts 2.8 million refugees).

During this period, government funding for the refugee determination process failed to keep pace. Although the Immigration and Refugee Board received a series of temporary funding injections, and although the Board repeatedly exceeded its funded output targets, intake persistently outstripped funded processing capacity, sometimes by wide margins. In the 2024-25 fiscal year, for example, the RPD received a funded finalization target of 85,000, which is less than half of its intake during that same period.[6] The result was mounting backlogs and lengthening wait times, with pending claims across the system growing from roughly 71,000 in 2018 to approximately 300,000 by the end of 2025.[7]

While political actors and the Canadian government expressed general concerns about the increasing number of claims and the growing backlog, much of their attention targeted specific groups of claimants. One such group was claimants who transited to Canada via the United States. The Safe Third Country Agreement between Canada and the United States ("the Agreement") allows Canada to return most people who come to a land border crossing and ask for refugee protection without adjudicating their claim.[8] However, the original Agreement did not apply to people crossing the border irregularly.[9] Following Donald Trump's first election to the presidency in 2016, increased numbers of refugee claimants used irregular crossings to make claims in Canada

---

[4] Petra Molnar Diop, "The 'Bogus' Refugee: Roma Asylum Claimants and Discourses of Fraud" (2014) 30:1 Refuge 67; John Carlaw, "Authoritarian Populism and Canada's Conservative Decade (2006–2015) in Citizenship and Immigration: The Politics and Practices of Kenneyism and Neo-conservative Multiculturalism" (2017) 51:3 Journal of Canadian Studies 782 at 800-803.

[5] United Nations High Commissioner for Refugees, Refugee Data Finder (2026), online: <https://www.unhcr.org/refugee-statistics/download/?v2url=e67958>.

[6] Immigration, Refugees and Citizenship Canada, "COW – Immigration and Refugee Board of Canada Main Estimates (2025–2026)" (9 June 2025), online: <https://www.canada.ca/en/immigration-refugees-citizenship/corporate/transparency/committees/cow-jun-9-2025/irb-main-estimates.html>.

[7] Immigration and Refugee Board, "Refugee Protection Claims (New System) by Country of Alleged Persecution - 2018", online: <https://www.irb-cisr.gc.ca/en/statistics/protection/Pages/RPDStat2018.aspx>; Immigration and Refugee Board, "Claims by Country of Alleged Persecution - 2025", online: <https://www.irb-cisr.gc.ca/en/statistics/protection/Pages/RPDStat2025.aspx>.

[8] Agreement Between the Government of Canada and the Government of the United States of America for Cooperation in the Examination of Refugee Status Claims from Nationals of Third Countries, 5 December 2002 (entered into force 29 December 2004) [Safe Third Country Agreement], implemented through the *Immigration and Refugee Protection Act*, SC 2001, c 27, ss 101(1)(e) & 102 [IRPA] and the *Immigration and Refugee Protection Regulations*, SOR/2002-227, ss 159.1–159.7 [IRPR]. For a discussion of the agreement, see Audrey Macklin, "Disappearing Refugees: Reflections on the Canada-US Safe Third Country Agreement" (2005) 36 Columbia Human Rights L Rev 365.

[9] Safe Third Country Agreement, ibid, art 4(1); IRPR, supra note 8, s 159.4.

– including a well-documented irregular crossing at Roxham Road on the Quebec–New York border.[10] This prompted substantial pushback from Conservative politicians, who inaccurately characterized those crossing at Roxham Road as "illegal" border crossers engaged in "queue jumping", and who demanded that the government "close the loophole" in the Agreement that left people crossing between ports of entry eligible to claim refugee protection in Canada.[11]

Ultimately, the Liberal government came to view itself as politically vulnerable to these critiques and pursued new arrangements with the US. After securing US permission to fully close the Canada-US border to asylum seekers during the pandemic – ostensibly to protect public health[12] – Canada persuaded the US to make the measure permanent: in March 2023, Canada and the United States entered into an Additional Protocol extending the Agreement across the entire land border.[13]

As critics predicted,[14] however, extending the Agreement and closing off safe routes into Canada offered by irregular land crossing at places like Roxham Road did not reduce the number of refugee claims made in Canada. While open irregular land crossings collapsed almost immediately after the Additional Protocol took effect,[15] claimants shifted to other routes, including arriving by air and making claims inland – resulting in the increased total number of claims noted above. This in turn prompted closer attention to claims made by people already in Canada, especially claims made by international students and other temporary residents.

[10] Craig Damian Smith, "Policy Change, Threat Perception, and Mobility Catalysts: The Trump Administration as Driver of Asylum Migration to Canada" (2023) 57:4 International Migration Rev 1678.
[11] See e.g. Teresa Wright, "Scheer calls on Trudeau to put a stop to 'queue jumping' at Canada-U.S. border", Global News (15 June 2018), online: <https://globalnews.ca/news/4278239/illegal-border-crossers-scheer-trudeau-letter>; Andy Blatchford, "Scheer vows to close asylum 'loophole', unveils immigration plan at Roxham Road", Global News (9 October 2019), online: <https://globalnews.ca/news/6010451/scheer-vows-to-close-asylum-loophole-unveils-immigration-plan-at-roxham-road-crossing>; Sean Boynton, "Poilievre calls on Trudeau to implement plan to close Roxham Road border crossing within 30 days", Global News (21 February 2023), online: <https://globalnews.ca/news/9501397/quebec-roxham-road-border-poilievre/>. The characterization of irregular border crossings as "illegal" is inaccurate for several reasons, including that Canada is prohibited under both international and domestic law from imposing penalties on asylum seekers who come to the country irregularly. See Audrey Macklin, "Citizenship, Non-Citizenship and the Rule of Law" (2018) 69 UNBLJ 19 at 31-32. And "queue jumping" is similarly inaccurate (and offensive), in part because there is no queue. See Johanna Reynolds & Jennifer Hyndman, "Shifting Grounds of Asylum in Canadian Public Discourse and Policy" in Vinh Nguyen & Thy Phu, eds, *Refugee States: Critical Refugee Studies in Canada* (Toronto: U Toronto P, 2021) 23 at 32.
[12] See Elise Mercier & Sean Rehaag, "The Right to Seek Asylum in Canada (During a Global Pandemic)" (2021) 57:3 Osgoode Hall LJ 705 at 719-722; Sean Rehaag, Janet Song & Alexander Toope, "Never Letting a Good Crisis Go to Waste: Canadian Interdiction of Asylum Seekers" (2020) 2 Frontiers in Human Dynamics 588961.
[13] Additional Protocol to the Agreement Between the Government of Canada and the Government of the United States of America for Cooperation in the Examination of Refugee Status Claims from Nationals of Third Countries (in force 25 March 2023), as implemented through IRPR, supra note 8, s 159.4(1.1). See also Efrat Arbel, "Bordering by sleight of hand: rights, race, and the unmaking of Canada's refugee protection system" (forthcoming) Int'l J Refugee Law, draft available online: <https://doi.org/10.1093/ijrl/eeag025>; Audrey Macklin, "Is the Canada-US Safe Third Country Agreement Doing What it is Supposed to Do?" (May 2026) 7 IRPP Policy, online: <https://centre.irpp.org/wp-content/uploads/sites/3/2026/05/Is-the-Canada-US-Safe-Third-Country-Agreement-doing-what-it-is-supposed-to-do-1.pdf>.
[14] See e.g. Jamie Liew, Petra Molnar & Julie Young, "The new US-Canada border deal is inhumane – and deadly", Al Jazeera (19 April 2023), online: <https://www.aljazeera.com/opinions/2023/4/19/the-new-us-canada-border-deal-is-inhumane-and>; Toula Drimonis, "You Close Roxham Road, People Are Just Going to Cross Somewhere Else", in The Walrus (19 March 2024), online: <https://thewalrus.ca/close-roxham-road>.
[15] Immigration and Refugee Board, "Irregular border crosser statistics", online: <https://www.irb-cisr.gc.ca/en/statistics/Pages/Irregular-border-crosser-statistics.aspx>.

The focus on students also had a broader context. By 2023, public and political debate had come to link high levels of temporary migration and of international students in particular to a worsening housing affordability crisis.[16] In January 2024, the government responded with a cap on new study permit applications, projected to cut approvals by 35 percent from the prior year, framed expressly as a measure to relieve pressure on housing.[17] Further restrictions followed, including narrowed eligibility for post-graduation work permits, a second cut to the study permit cap, and a revised Immigration Levels Plan that reduced permanent resident targets and, for the first time, set targets for temporary residents.[18] Having first recruited international students in record numbers and then abruptly narrowed their options for remaining in Canada, the government confronted the inevitable result: some of those left without a pathway to status turned to the refugee system. In government messaging, such claims were increasingly framed as misuse of asylum rather than as a foreseeable consequence of a policy reversal.[19]

These pressures also coincided with a global resurgence of far-right, racist, anti-immigrant politics. During the same period, US politics, which are followed carefully by many in Canada, had become saturated with anti-immigrant rhetoric. Donald Trump's first presidential campaign in 2015 included repeated descriptions of Mexican migrants as criminals and rapists,[20] and anti-immigrant appeals – the border wall, the Muslim ban, family separation – remained central to his first term.[21]

---

[16] Félix Blain, et al, "Feasibility study: Estimating the international student population in Canada using administrative data" (2026) Statistics Canada Report (81-595-M2026001), online: <https://www150.statcan.gc.ca/n1/en/catalogue/81-595-M2026001>; Marieke Walsh, "Federal government should look at cap on student visas, Housing Minister Sean Fraser says: Number of international students in Canada has more than doubled since Trudeau took office in 2015, government data show" in The Globe and Mail (21 August 2023); Environics Institute for Survey Research, "Canadian Public Opinion About Immigration and Refugees – Fall 2023" (2023), online: <https://environicsinstitute.org/public-opinion-about-immigration-refugees/> at 8 (finding that housing costs had become the most commonly cited concern among those who believe there is too much immigration).

[17] Immigration, Refugees and Citizenship Canada, News Release, "Canada to stabilize growth and decrease number of new international student permits issued to approximately 360,000 for 2024" (22 January 2024), online: <https://www.canada.ca/en/immigration-refugees-citizenship/news/2024/01/canada-to-stabilize-growth-and-decrease-number-of-new-international-student-permits-issued-to-approximately-360000-for-2024.html>.

[18] Immigration, Refugees and Citizenship Canada, "2025–2027 Immigration Levels Plan" (2024), online: <https://www.canada.ca/en/immigration-refugees-citizenship/news/2024/10/20252027-immigration-levels-plan.html>.

[19] See e.g. Sean Boynton, "'Alarming trend' of more international students claiming asylum: minister", Global News (22 September 2024), online: <https://globalnews.ca/news/10766777/immigration-international-students-asylum-miller-west-block>.

[20] Laura Finley & Luigi Esposito, "The Immigrant as Bogeyman: Examining Donald Trump and the Right's Anti-immigrant, Anti-PC Rhetoric" (2020) 44:2 Humanity & Society 178 at 187-189.

[21] Sophia Jordán Wallace & Chris Zepeda-Millán, *Walls, Cages, and Family Separation: Race and Immigration Policy in the Trump Era* (Cambridge, UK: Cambridge University Press, 2020).

The rhetoric further escalated in the 2024 campaign, in which Trump promised the largest deportation operation in American history[22] and repeatedly said that immigrants were "poisoning the blood" of the country.[23] Across Europe and beyond, 2024 delivered gains for political parties campaigning on hostility to migration, and, as those parties moved from the fringes toward power, mainstream parties responded by adopting increasingly restrictive asylum and immigration policies of their own.[24]

Canada was not immune to this trend. After years of politicians blaming immigrants and temporary migrants for affordability concerns, in 2024, for the first time in a quarter century, a majority of Canadians told the Environics Institute's long-running Focus Canada survey that there is too much immigration to Canada,[25] and a growing proportion agreed that "many people claiming to be refugees are not real refugees".[26] While anti-immigrant policies and rhetoric have long histories in Canadian politics,[27] recent years have seen a marked escalation,[28] and the two leading federal parties converged on more restrictionist positions.[29]

Then came the return of Donald Trump to the presidency in January 2025. Within hours of taking office, the new administration issued a barrage of executive actions involving immigration enforcement: declaring a border emergency, suspending refugee resettlement and asylum processing, stripping protections from humanitarian parolees, and launching a mass deportation campaign that saw arrests, detentions, and removals multiply.[30] Particularly difficult for the Canadian government was that the new US administration justified sweeping new tariffs against

---

[22] Joel Rose & Sergio Martínez-Beltrán, "Trump touts historic deportation plans, but his own record reveals big obstacles" in NPR (14 August 2024), online: <https://www.npr.org/2024/08/14/nx-s1-5037992/trump-immigrants-border-mass-deportation-presidential-race-migrants>.
[23] Ginger Gibson, "Trump says immigrants are 'poisoning the blood of our country.' Biden campaign likens comments to Hitler" in NBC News (17 December 2023), online: <https://www.nbcnews.com/politics/2024-election/trump-says-immigrants-are-poisoning-blood-country-biden-campaign-liken-rcna130141>.
[24] Cas Mudde, "The Far Right and the 2024 European Elections" (2024) 59:2 Intereconomics 61.
[25] Environics Institute for Survey Research, "Canadian Public Opinion About Immigration and Refugees – 2024 Series" (2024), online: <https://environicsinstitute.org/canadian-public-opinion-about-immigration-and-refugees-2024/> at 6 (finding that, "for the first time in a quarter century, a clear majority of Canadians say there is too much immigration").
[26] Ibid at 10 (reporting that 43 percent agreed, up 7 points from 2023).
[27] Laura Kwak, "Problematizing Canadian exceptionalism: A study of right-populism, white nationalism and Conservative political parties" (2020) 10:6 Oñati Socio-legal Series 1166.
[28] Rupa Banerjee, et al, "Canada's Long-Standing Openness to Immigration Comes Under Pressure" in *Migration Policy Institute* (24 June 2025), online: <https://www.migrationpolicy.org/journal/country-profile/canadas-long-standing-openness-immigration-comes-under-pressure>; Jillian Kestler-D'Amours, "Inside Canada's 'troubling' shift on migrant, refugee rights" in Al Jazeera (17 June 2026), online: <https://www.aljazeera.com/news/2026/6/17/inside-canadas-troubling-shift-on-migrant-refugee-rights>.
[29] Ibid (Banerjee); John Tasker, "Poilievre calls for 'very hard caps' on immigration to better integrate newcomers" in *CBC News* (14 July 2025), online: <https://www.cbc.ca/news/politics/poilievre-immigraton-very-hard-cap-1.7584743>.
[30] See Muzaffar Chishti, "Unleashing Power in New Ways: Immigration in the First Year of Trump 2.0" in Migration Policy Institute (13 January 2026), online: <https://www.migrationpolicy.org/journal/policy-beat/unleashing-power-new-ways-immigration-first-year-trump-20>.

Canada partly by reference to migrants said to be flowing across the northern border,[31] while musing publicly about annexation.[32] Not surprisingly, partly in response to these US threats, border security moved to the centre of Canadian politics.[33]

It was against this backdrop – a mounting backlog, scapegoating of temporary foreign residents (and especially international students) for housing affordability, and border politics inflamed from the south – that the one-year bar made its legislative debut.

### *2.2 From Bill C-2 to Bill C-12*

The one-year bar first appeared on June 3, 2025, in Bill C-2, the Strong Borders Act: a sprawling omnibus bill covering border security, police powers, information sharing, money laundering, and immigration, introduced as the first major legislation of the new Liberal government under Prime Minister Mark Carney.[34]

Critical responses were swift. Hundreds of civil society organizations signed joint statements calling for withdrawal of the bill.[35] Much of the civil society opposition to the bill focused not on the refugee provisions but on its surveillance measures, including powers permitting warrantless government demands for subscriber information from a wide range of service providers, and cross-border data-sharing provisions designed to align Canadian law with the American CLOUD Act and the Second Additional Protocol to the Budapest Convention on Cybercrime.[36]

In October 2025 the government carved the immigration measures out of the omnibus and reintroduced them as Bill C-12, the Strengthening Canada's Immigration System and Borders Act.[37] From there, things moved quickly: second reading in the House of Commons on October 23, committee study in November, third reading on December 11, Senate passage on March 12, 2026. The Senate's social affairs committee took the unusual step of recommending amendments, including deletion of the one-year bar.[38] The government declined, and the Senate did not insist. Royal assent came on March 26, 2026.[39]

---

[31] White House, "Imposing Duties to Address the Flow of Illicit Drugs Across Our Northern Border" (1 February 2025), online: <https://www.whitehouse.gov/presidential-actions/2025/02/imposing-duties-to-address-the-flow-of-illicit-drugs-across-our-national-border>.
[32] Murray Brewster, "Trump is starting a trade war. If he wants to absorb Canada, what comes next will be worse" in CBC News (1 February 2025), online: <https://www.cbc.ca/news/politics/trump-trade-war-vs-economic-war-1.7447927>.
[33] Public Safety Canada, News Release, "Government of Canada expands plan to strengthen border security" (4 February 2025), online: <https://www.canada.ca/en/public-safety-canada/news/2025/02/government-of-canada-expands-plan-to-strengthen-border-security.html>; Catharine Tunney, "Liberals hoped their border bill would quickly pass. Now they're aiming for next year", CBC News (22 November 2025), online: <https://www.cbc.ca/news/politics/border-security-bill-c-12-9.6988286>.
[34] Bill C-2, *Strong Borders Act*, 1st Sess, 45th Parl, 2025 (first reading 3 June 2025).
[35] International Civil Liberties Monitoring Group, "Joint Statement on Bill C-12" (9 October 2025), online: <https://iclmg.ca/joint-statement-bill-c-12>.
[36] See e.g. Kate Robertson, "Unspoken Implications: A Preliminary Analysis of Bill C-2 and Canada's Potential Data-Sharing Obligations Towards the United States and Other Countries", Citizen Lab Report No 187 (16 June 2025), online: <https://citizenlab.ca/2025/06/a-preliminary-analysis-of-bill-c-2>.
[37] Bill C-12, *Strengthening Canada's Immigration System and Borders Act*, 1st Sess, 45th Parl, 2025 (assented to 26 March 2026) [Bill C-12] (first reading in the House of Commons, 8 October 2025).

### *2.3 The one-year bar and the PRRA*

The one-year bar is set out in new paragraph 101(1)(b.1) of the IRPA.[40] Under the new provision, a claim is ineligible for referral to the RPD if the claimant entered Canada after June 24, 2020, and waited more than one year after entry to apply for refugee protection. Where there have been multiple entries, time runs from the day after the first one. Leaving Canada and returning does not reset the clock.[41]

The new provision does not contain any exceptions or allow claimants an opportunity to explain why they delayed in claiming. There is no allowance for newly arisen risks, changed conditions in the country of origin, trauma, lack of information, or other possible explanations. Nor has the regulatory process, which can carve out exceptions to the one-year bar,[42] been deployed to address these concerns. In June 2026, the government published proposed amendments to the Immigration and Refugee Protection Regulations for consultation.[43] The proposed regulations create exceptions to the one-year bar in two circumstances: unaccompanied minors[44] and people who submitted documents required to make refugee claims within one year, even if they were not able to formally make a refugee claim due to delays in scheduling meetings with immigration officers.[45]

It is important to recognize that a person caught by the one-year bar is not left without any process. They can seek a PRRA, which assesses the same risks that ground refugee protection: persecution in the sense of the Refugee Convention, as well as torture, risk to life, or cruel and unusual treatment.[46] On paper, then, the substantive tests for refugee protection are unchanged. However, procedure is another matter.

---

[38] David Baxter, "Senate committee recommends removing broad immigration powers from border bill", CBC News (23 February 2026), online: <https://www.cbc.ca/news/politics/senate-refugee-claims-c12-9.7103017>; Canada, Parliament, Senate, Standing Senate Committee on Social Affairs, Science and Technology, *Sixth Report: Subject Matter of Parts 5, 6, 7 and 8 of Bill C-12, An Act respecting certain measures relating to the security of Canada's borders and the integrity of the Canadian immigration system and respecting other related security measures*, 45-1 (20 February 2026) [Senate Committee Report].
[39] Bill C-12, supra note 37.
[40] IRPA, supra note 8, as amended by Bill C-12, supra note 37.
[41] Ibid, s 101(1.1).
[42] Ibid, s 111.1(1)(b.1).
[43] Regulations Amending the Immigration and Refugee Protection Regulations (Asylum System Reform), (2026) C Gaz I, vol 160, no 25, 1332 (proposed regulations for consultation).
[44] Ibid at 1364 (amending IRPR, supra note 8, s 159.002).
[45] Ibid at 1364 (amending IRPR, supra note 8, s 159.001).
[46] IRPA, supra note 8, ss 96-97 & 112-113.

A PRRA is decided by an immigration officer, typically evaluated solely on a written record, within tight timelines (15 days to submit the application after notification, and a further 15 for evidence and submissions). An oral hearing is available only where prescribed factors are met, principally where credibility is determinative, and there are limits on what issues can be addressed at oral hearings.[47] Parliament has taken a different approach to one other group of ineligible claimants who only get access to PRRAs. Where a claim is ineligible solely because the claimant previously sought asylum in a country with which Canada has an information-sharing agreement, a hearing must be held unless the application can be allowed without one.[48] Bill C-12 did not extend that requirement to claimants caught by the one-year bar.

This contrasts with the usual full refugee determination process. At the RPD, an oral hearing is the default, a design choice that traces back to *Singh v Minister of Employment and Immigration* ("*Singh*"),[49] where the Supreme Court held that section 7 of the *Canadian Charter of Rights and Freedoms* ("Charter")[50] requires an oral hearing where a refugee claim turns on credibility.[51] As a result of this decision, the IRPA explicitly obliges the RPD to hold a hearing before rejecting a claim.[52] A refused claimant can also usually appeal to the RAD, which reviews the case on the merits on a mostly de novo (i.e. non-deferential) standard, can receive new evidence, and whose proceedings automatically delay the coming into force of removal orders[53] – and the RAD's decision can then be the subject of an application for leave and judicial review in the Federal Court, during which time the claimant again benefits from an automatic stay.[54]

For a refused PRRA, on the other hand, the only recourse is an application for leave and judicial review in the Federal Court on a deferential standard of review, with no automatic stay – meaning that removal can proceed unless a judge grants a stay.[55] In practice, then, the PRRA will not only replace first-instance refugee adjudication but, for many claimants, also be the final determination.

Beyond the question of recourse for refused applications, there is also a transparency gap.

Refugee decision-making at Canada's Immigration and Refugee Board has been heavily scrutinized through decades of legal research, facilitated in part by the publication of tens of thousands of refugee decisions, by the Board's openness to engaging with researchers, and by the Board's timely compliance with access to information requests.[56] PRRA decisions, by contrast, are not published, the government appears more reluctant to engage with critical researchers than the IRB, and obtaining granular and timely data about PRRA decision-making through access to information requests is extremely challenging. The direct result is that while there are dozens of published studies about IRB refugee decision-making, there are only a handful focusing on PRRAs – and the only existing large-scale empirical analysis of PRRA decision-making relied exclusively on Federal Court judicial reviews.[57] Indeed, the lack of transparency extends beyond scholarship: the Senate committee that recommended removing the one-year bar from Bill C-12 expressed frustration that government "officials have not provided all of the requested data to support their claims".[58]

---

[47] Ibid, ss 112-114; IRPR, supra note 8, ss 160-162 & 167-168.
[48] IRPA, supra note 8, ss 101(1)(c.1) & 113.01.
[49] *Singh v Minister of Employment and Immigration*, [1985] 1 SCR 177 [*Singh*].
[50] *Canadian Charter of Rights and Freedoms*, Part I of the *Constitution Act*, 1982, being Schedule B to the *Canada Act 1982* (UK), 1982, c 11.
[51] *Singh*, supra note 49 at para 59.
[52] IRPA, supra note 8, s 170(b).
[53] Ibid, ss 49 & 110-111. On the non-deferential standard of review, see *Canada (Citizenship and Immigration) v Huruglica*, 2016 FCA 93 at para 78.
[54] IRPA, supra note 8, s 72; IRPR, supra note 8, s 231.
[55] IRPA, supra note 8, s 72; IRPR, supra note 8, s 231. On the deferential standard of review, see *Canada (Minister of Citizenship and Immigration) v Vavilov*, 2019 SCC 65 at paras 83-86.

Despite the transparency gap, some patterns are evident in the record. Based on high-level statistics, for example, it seems that outcomes in the two processes differ significantly. In its brief to Parliament on the bill, the Canadian Council for Refugees reported a PRRA acceptance rate of 33 percent, against a five-year average acceptance rate at the IRB of 63 percent.[59] The comparison is imperfect because the two caseloads differ, but the overall pattern is unsurprising. The PRRA was designed as a residual safety net for people whose claims had already been heard and refused, a final check against refoulement before removal mostly to account for changed circumstances since the time of a refugee hearing.

Two further features of the one-year bar are worth noting. First, the bar applies retroactively, to claims made on or after June 3, 2025 (the date Bill C-2 was tabled), even though the measure did not become law for almost ten months afterward.[60] And, second, the affected population is large. Shortly after royal assent, the immigration department began sending procedural fairness letters to

---

[56] See e.g. Sean Rehaag, "Claim Types in Canada's Refugee Determination System: An Empirical Snapshot (2013-2021)" (2024) 40:1 Refuge 1; Efrat Arbel, "The Culture of Rights Protection in Canadian Refugee Law: Examining the Domestic Violence Cases" (2013) 58:3 McGill LJ 729; Catherine Dauvergne, "International Human Rights in Canadian Immigration Law – The Case of the Immigration and Refugee Board of Canada" (2012) 19:1 Indiana J of Global Legal Studies 305; Nicole LaViolette, "Independent Human Rights Documentation and Sexual Minorities: An Ongoing Challenge for the Canadian Refugee Determination Process" (2009) 13:2-3 Int'l J of Human Rights 437; Sean Rehaag, "Troubling Patterns in Canadian Refugee Adjudication" (2008) 39:2 Ottawa L Rev 335; Audrey Macklin, "Cross-Border Shopping for Ideas: A Critical Review of United States, Canadian, and Australian Approaches to Gender-Related Asylum Claims" (1998) 13 Georgetown Immigration Law Journal 25.

[57] Simon Wallace, "Getting it Right the First Time: Exploring the False Economy of Bill C-12's Refugee Process Shortcuts", Refugee Law Lab Working Paper (17 October 2025), online: <https://ssrn.com/abstract=5620250>.

[58] Senate Committee Report, supra note 38.

[59] Canadian Council for Refugees, "Bill C-12: Strengthening Canada's Immigration System and Borders Act: Submission to the Standing Committee on Public Safety and National Security (SECU)" (November 2025), online: <https://ccrweb.ca/sites/ccrweb.ca/files/2026-01/CCR_C-12%20Brief_SECU.pdf>.

[60] Bill C-12, supra note 37.

people whose pending claims were now presumptively ineligible, a group estimated at approximately 30,000.[61]

### *2.4 The government's case*

The case the government has made for the one-year bar has three main arguments.

The first main argument is about efficiency and responding to the large inventory of claims at the IRB. In announcing the coming into force of Bill C-12, the government indicated that the time limit will "reduce pressure on the asylum system".[62] Along similar lines, Minister Metlege Diab told a Senate committee that "[w]e have a large number of applications in the queue. Processing times are long, and our system is not well equipped to handle further increases. […] These measures help prevent backlogs and allow the Immigration and Refugee Board to focus on the claims before it."[63]

The second main argument is about incentives. The government has pointed to claims by people whose temporary status is running out, and by international students in particular, as evidence that the asylum system was being used as a "shortcut to regular immigration pathways".[64] Minister Metlege Diab put the point more bluntly in the House of Commons, describing Bill C-12 as legislation that would "curb the misuse of asylum".[65] On this view, a firm deadline will incentivize people with genuine fears to claim in a timely manner, and will deter people from using the system strategically "to avoid other immigration processes or extend their stay in Canada".[66] As the government put it: "The one-year time limit was chosen in part because a significant majority of claimants make their claim within their first year of being in Canada… The use of a one-year period is […] a clear and easily communicated timeline for claimants to understand the period in which they are allowed to make a claim."[67]

---

[61] Raffy Boudjikanian, "Immigration Department informing some 30,000 applicants they may be ineligible for refugee hearings", CBC News (10 April 2026), online: <https://www.cbc.ca/news/politics/immigration-thirty-thousand-ineligible-asylum-9.7158489>.

[62] Government of Canada, "New immigration and asylum measures from Bill C-12 (the Strengthening Canada's Immigration System and Borders Act) have become law" (26 March 2026), online: <https://www.canada.ca/en/immigration-refugees-citizenship/news/2026/03/new-immigration-and-asylum-measures-from-bill-c-12-the-strengthening-canadas-immigration-system-and-borders-act-have-become-law.html>.

[63] Canada, Senate, Standing Senate Committee on Social Affairs, Science and Technology, *Evidence*, 45-1, No 25 (12 February 2026) at 25:3 (Hon Lena Metlege Diab) [Senate Evidence No 25].

[64] Ibid.

[65] *House of Commons Debates*, 45-1, No 87 (23 February 2026) at 1415 (Hon Lena Metlege Diab).

[66] Senate Evidence No 25, supra note 63 at 25:23 (Jason Hollmann, Director General, Asylum, IRCC).

[67] Raffy Boudjikanian, "Asylum crackdown passed under Carney first requested by Trudeau's immigration minister" in CBC News (15 April 2026), online: <https://www.cbc.ca/news/politics/immigration-minister-asylum-crackdown-trudeau-carney-9.7163661>.

The third main argument is about adequacy and the contention that nothing of substance is lost when claimants are directed to PRRAs instead of the regular system. Minister Metlege Diab told senators that the change is actually "better for the applicants", on the theory that clearly meritorious cases will be granted on paper without waiting years for a hearing.[68] Pressed on *Singh* and the constitutional requirement for an oral hearing, departmental counsel responded that the Supreme Court required an oral hearing only where credibility is in issue, and that the PRRA regulations already provide for hearings in those cases.[69] The Minister of Justice's Charter Statement about the bill took the same line: because the PRRA assesses the same risks the Board would have assessed, the diversion is said to be consistent with section 7 of the Charter.[70]

Notice how each of these arguments involves empirically testable presumptions. Efficiency assumes the diverted claims would otherwise have consumed Board resources without much changing outcomes. Incentivization assumes that late claims are disproportionately strategic and not well-founded. Adequacy assumes that the PRRA process, without an administrative appeal, can reliably identify meritorious claims. And beneath these arguments is the basic premise that one year is enough time for refugees to come forward.

### *2.5 The critics' case*

The critics' response begins with the Refugee Convention, which defines a refugee by reference to a well-founded fear of persecution, not by how quickly protection was sought.[71] As the Canadian Council for Refugees told a Senate committee, "[t]he UN convention puts no timelines on your right to asylum."[72]

The critics do not, of course, deny that delay can matter in refugee adjudication. To the contrary, case law clearly establishes that delay can impact credibility and subjective fear, and the Board has long weighed it that way.[73] Rather, the objection is to what the one-year bar does with delay.

[68] Senate Evidence No 25, supra note 63 at 25:26 (Hon Lena Metlege Diab).
[69] Senate Evidence No 25, supra note 63 at 25:21 (Joanie Roy-Craswell, Acting Senior Counsel, Legal Service, IRCC).
[70] Canada, Department of Justice, "Charter Statement – Bill C-12: An Act respecting certain measures relating to the security of Canada's borders and the integrity of the Canadian immigration system and respecting other related security measures" (tabled 5 November 2025), online: <https://www.justice.gc.ca/eng/csj-sjc/pl/charter-charte/c12_3.html> [Charter Statement].
[71] Convention Relating to the Status of Refugees, 28 July 1951, 189 UNTS 137, art 1A(2). See also IRPA, supra note 8, ss 96 & 97.
[72] Canada, Senate, Standing Senate Committee on Social Affairs, Science and Technology, *Evidence*, 45-1, No 24 (10 February 2026) at 24:25 (Gauri Sreenivasan) [Senate Evidence No 24].
[73] See e.g. James Hathaway & Michelle Foster, *The Law of Refugee Status, 2nd ed* (Cambridge: Cambridge UP, 2014) at 98-99; *Huerta v Canada (Minister of Employment and Immigration)* (1993), 157 NR 12 (FCA) at para 4; *Zhou v Canada (Citizenship and Immigration)*, 2020 FC 676 at para 24.

Delay was previously assessed case by case, together with the claimant's explanation for it.[74] The bar converts that individualized, context-sensitive exercise into an absolute eligibility rule that admits of no explanation. It was this conversion that led the Canadian Bar Association's (CBA) immigration law section to call the one-year rule "arbitrary and without legal justification" and to urge that it be withdrawn.[75] As the CBA put it, "[d]ecades of jurisprudence have emphasized the need for individualized assessments in refugee determinations and established that delay in advancing a claim should not, in itself, unequivocally undermine one's ability to receive a Charter-mandated hearing."[76]

The critics also note that this arbitrariness has the potential to impact different groups of claimants differently because the people who need more than a year to make a claim may not be a random cross-section of claimants. When the CBA and Amnesty International warned senators that the changes build a two-tier asylum system, they emphasized that the lower tier – assessment without a guaranteed hearing and with limited procedural rights – will be disproportionately populated by vulnerable groups, including SOGIE and gender-based claimants.[77] The Canadian Civil Liberties Association made a similar point: survivors of torture, gender-based violence, and anti-SOGIE persecution may need years before they "feel safe enough to disclose their identity or share their experiences".[78]

Nor, as the critics point out, is delayed disclosure the only mechanism by which a well-founded claim can surface late. Sometimes the risk itself arises only after arrival – in other words, sur place claims.[79] The Canadian Association of Refugee Lawyers (CARL), for example, noted that people claim after a year "for fully legitimate reasons", including a change of government back home, an outbreak of war, or human rights activism in Canada that makes them a target.[80]

---

[74] Ibid (Hathaway & Foster) at 99; Immigration and Refugee Board, "Assessment of Credibility" (31 December 2021), online: <https://www.irb-cisr.gc.ca/en/legal-policy/legal-concepts/Pages/Credib.aspx> at 2-45 – 2-60; *X (Re)*, 2021 CanLII 155387 (21 June 2021) at 12-15 (allowing the appeal of a domestic violence survivor whose over three-year delay had been held against her subjective fear); *X (Re)*, RAD File TC0-00944 (26 April 2021) at 10-12 (allowing the appeal of a gay claimant from India whose four-month delay in claiming after arrival in Canada had grounded an adverse credibility inference).
[75] Canadian Bar Association, Immigration Law Section, Submission on Bill C-12 (Strengthening Canada's Immigration System and Borders Act) (4 November 2025), online: <https://cba.org/Our-Impact/Submissions/Bill-C-12-Strengthening-Canada-s-Immigration-System-and-Borders-Act> at 2.
[76] Ibid.
[77] Canadian Bar Association, "Open letter to all senators re: Bill C-12" (24 February 2026), online: <https://cba.org/Our-Impact/Submissions/Letter-to-Senators-about-Bill-C-12> at 2; Senate Evidence No 24, supra note 72 at 24:9 (Julia Sande).
[78] Canadian Civil Liberties Association, "Submission to the House of Commons Standing Committee on Public Safety and National Security (SECU)" (12 November 2025), online: <https://ccla.org/wp-content/uploads/2025/11/2025.10.12-CCLA_SECU_Brief_C-12-1.pdf> at 5.
[79] Hathaway & Foster, supra note 73 at 75-90.
[80] Raffy Boudjikanian, "Immigration minister defends sweeping new powers in border bill" in *CBC News* (5 June 2025), online: <https://www.cbc.ca/news/politics/liberal-border-bill-asylum-immigration-1.7553227>.

Among the bill's critics, the late disclosure problem has been raised as a particular concern for sexual minority claims. Rainbow Railroad, a leading organization working with SOGIE refugees in Canada, told a House of Commons committee that disclosure of sexual orientation or gender identity typically comes only once safety and trust are established; that some claimants – young people especially – may come to understand or name their identity only years after arrival; and that a rigid deadline "institutionalizes disbelief in queer and trans flight narratives".[81]

Canada's refugee determination system explicitly agrees with this framing. The Immigration and Refugee Board's Guideline 9 instructs decision-makers that delayed disclosure of sexual orientation or gender identity should not, without more, ground an adverse credibility finding, because it may be difficult to disclose sexual orientation at early stages in the process, because claimants may reasonably fear reprisals for themselves or others, because claimants may be hesitant to share their identity with family members, or because it may take time for claimants to develop or embrace their identity.[82] Guideline 4 makes similar points about gender-based violence and sexual violence, characterizing expectations that survivors will report, disclose, or otherwise seek assistance in a timely fashion as being based on "myths, stereotypes, and incorrect assumptions".[83]

Academic research also backs up this framing. Scholarship in several jurisdictions has documented that sexual minority claimants frequently claim late or disclose late – and are too often unreasonably penalized for it in credibility assessment.[84] Research has also documented that transgender claimants frequently live in Canada on various visa statuses for some time before launching their claims.[85] And, since the 1990s, scholarship in Canada and elsewhere has come to similar conclusions about delays in making gender-based violence refugee claims.[86]

---

[81] Rainbow Railroad, "Brief Submitted to the House of Commons Standing Committee on Citizenship and Immigration regarding Bill C-12" (November 2025), online: <https://www.ourcommons.ca/Content/Committee/451/CIMM/Brief/BR13730168/br-external/RainbowRailroad-e.pdf> at 3-4.

[82] Immigration and Refugee Board, "Chairperson's Guideline 9: Proceedings Before the IRB Involving Sexual Orientation, Gender Identity and Expression, and Sex Characteristics" (17 December 2021), online: <https://www.irb-cisr.gc.ca/en/legal-policy/policies/Pages/GuideDir09.aspx> at 6, 7.4, 8.5.11 & 8.5.12.

[83] Immigration and Refugee Board, "Chairperson's Guideline 4: Gender Considerations in Proceedings Before the Immigration and Refugee Board" (31 October 2023), online: <https://www.irb-cisr.gc.ca/en/legal-policy/policies/Pages/GuideDir04.aspx> at 4.4.

[84] See e.g. Nuno Ferreira, "Better Late Than Never? SOGI Asylum Claims and Late Disclosure Through a Foucauldian Lens" (2023) 27:1 UCLA J Int'l L Foreign Aff 17; Sabine Jansen & Thomas Spijkerboer, *Fleeing Homophobia: Asylum Claims Related to Sexual Orientation and Gender Identity in Europe* (Amsterdam: COC Nederland & VU University Amsterdam, 2011), online: <https://www.refworld.org/reference/themreport/vuu/2011/83496> at 65-69.

[85] Alex Verman & Sean Rehaag, "Transgender Erasure: Barriers Facing Transgender Refugees in Canada" (2024) 69:1 McGill LJ 49 at 90.

[86] See e.g. Philip G Schrag, et al, "Rejecting Refugees: Homeland Security's Administration of the One-Year Bar to Asylum" (2010) 52:3 Wm & Mary L Rev 651 at 702; Constance MacIntosh, "Domestic Violence and Gender-Based

In addition to these substantive critiques, some critics also contest the bar on the government's own terms. On efficiency, for example, Simon Wallace – drawing on a dataset of over 180,000 Federal Court proceedings – has argued that the procedural changes represent a false economy. He contends that paper-based PRRA first-instance decisions will generate more errors and more litigation, shifting the backlog from the Board to the (even more expensive) Federal Court rather than eliminating it.[87] Research on a prior attempt by the former Conservative government to strip procedural rights from refugees, the 2012 designated country of origin reforms, points the same way, finding that stripping procedural rights did not make the system faster.[88]

As the bill moved through Parliament, critics offered several recommendations.

CARL, for example, proposed that, if a one-year bar is kept at all, it should work differently. The clock should start at the claimant's most recent entry into Canada, not their first, and it should reset when conditions in their home country change. Some groups should also be exempt, including people fleeing gender-based violence and other vulnerable claimants. And anyone diverted from the refugee process should be guaranteed a hearing and an appeal, with removal paused while the appeal and subsequent judicial review are decided.[89]

The UNHCR's recommendation to senators converged on the same core safeguards: anyone found ineligible under the new provisions should receive a mandatory hearing and an appeal to the Refugee Appeal Division.[90]

The Senate's social affairs committee, as noted above, went further still, recommending that the one-year bar be deleted from the bill altogether.[91]

Shortly before the bill received royal assent, the UN Human Rights Committee noted that it had received reports that the bill "may weaken refugee protection"[92] and recommended that Canada:

---

Persecution: How Refugee Adjudicators Judge Women Seeking Refuge from Spousal Violence—and Why Reform Is Needed" (2009) 26:2 Refuge 147 at 152; Diana Bögner, Jane Herlihy & Chris R Brewin, "Impact of Sexual Violence on Disclosure during Home Office Interviews" (2007) 191:1 Brit J Psychiatry 75 at 80; F Stairs & L Pope, "No Place Like Home: Assaulted Migrant Women's Claims to Refugee Status" (1990) 6 J L & Soc Pol'y 148 at 200.

[87] Wallace, supra note 57 at 33-34.

[88] Idil Atak, Graham Hudson & Delphine Nakache, "The Securitisation of Canada's Refugee System: Reviewing the Unintended Consequences of the 2012 Reform" (2018) 37:1 Refugee Survey Quarterly 1 at 24.

[89] Canadian Association of Refugee Lawyers, "Brief to the Standing Senate Committee on Social Affairs, Science and Technology regarding Bill C-12" (9 February 2026), online: <https://sencanada.ca/Content/Sen/Committee/451/SOCI/briefs/2026-02-09_SOCI_SM-C-12_Brief_CARL_e.pdf> at 4-5.

[90] Senate Evidence No 24, supra note 72 at 24:7-8 (Azadeh Tamjeedi, UNHCR).

[91] Senate Committee Report, supra note 38.

[92] UN Human Rights Committee, *Concluding Observations on the Seventh Periodic Report of Canada*, UNHRCOR, 146th Sess, UN Doc CCPR/C/CAN/CO/7 (2026).

> ensure that all persons seeking international protection have unfettered access to the national territory and to fair and efficient procedures, with all necessary procedural safeguards, for the individualized determination of refugee status or other forms of international protection, in full compliance with the principle of non-refoulement.[93]

None of these recommendations was taken up.

### *2.6 The American experience with the one-year bar*

Canada's one-year bar shares many similarities with an asylum rule in the United States. In 1996, Congress enacted a one-year filing deadline for asylum. Under that rule, an applicant must demonstrate "by clear and convincing evidence" that the application was filed within one year of arrival in the United States.[94] The American rule is less strict than the Canadian one in two respects. The clock runs from the applicant's most recent arrival, and the statute contains exceptions for changed circumstances that materially affect eligibility, and for extraordinary circumstances relating to the delay.[95]

Even with those safeguards, the record is sobering.

The leading empirical study analyzed a government database of over 380,000 asylum cases.[96] Between April 1998, when the deadline took effect, and June 2009, the Department of Homeland Security rejected 54,141 late applicants, approximately one in six applicants adjudicated in the period.[97] The authors estimated that, but for the deadline, adjudicators would likely have granted asylum in roughly 16,000 additional cases, covering more than 21,000 refugees once family members are counted.[98]

The study also found that the burdens of the US one-year bar were not evenly distributed. Women in the dataset filed late at a rate 13 percent higher than men (32.7 percent against 29 percent), and filed very late, four or more years after entry, at a rate over 50 percent higher than men.[99] Other researchers have come to similar conclusions, noting that the US one-year bar disproportionately impacts women facing gender-based violence and people facing persecution on account of their sexual orientation or gender identity.[100]

---

93 Ibid.
94 *Immigration and Nationality Act*, § 208(a)(2)(B), 8 USC § 1158(a)(2)(B), enacted by *Illegal Immigration Reform and Immigrant Responsibility Act of 1996*, Pub L No 104-208, div C, 110 Stat 3009-546 at 691 [*INA*].
95 8 USC § 1158(a)(2)(D).
96 Schrag, et al, supra note 86 at 681. See also Philip Schrag, Andrew Schoenholtz & Jaya Ramji-Nogales, *Lives in the Balance: Asylum Adjudication by the Department of Homeland Security* (New York: NYU Press, 2014) at 49ff.
97 Schrag, et al, supra note 86 at 683.
98 Ibid at 659.
99 Ibid at 702.

One further difference between the regimes deserves emphasis: an applicant in the US who misses the deadline still receives a hearing before an immigration judge on withholding of removal and protection under the Convention Against Torture. Applicants in the US lose access to full asylum status and its benefits, not to adjudication itself.[101] A Canadian claimant caught by the new one-year bar, by contrast, loses tribunal adjudication altogether. The Canadian statute also contains no exceptions, and its clock runs from first entry.

### *2.7 The missing Canadian evidence*

The Canadian debate over the one-year bar turns on empirical claims, but little Canadian evidence has been brought to bear on them. The government's premises, that genuine claimants claim within a year and that late claims typically involve misuse, were asserted without supporting analysis. The critics stand on firmer ground: the Board's own guidelines, qualitative scholarship, and the American record all suggest that late claims skew toward gender-based and SOGIE persecution. But that evidence is indirect. No published Canadian study measures how common late claims are, who makes them, or how they fare once decided.

The two accounts predict different things. If the government is right, late claims should be few and should seldom succeed. If the critics are right, a substantial number of late claims should be found well-founded, and two further patterns should appear. First, gender and SOGIE claims should be overrepresented among late claims. Second, among late claims, refusals of gender and SOGIE claims should be overturned more often than refusals of other claims, whether because these claims are late for legitimate reasons and so are stronger on average, or because, as the guidelines and the scholarship warn, adjudicators too readily hold delay against them.

These patterns matter because of how the one-year bar works, including removing the oral hearing, the appeal, and the automatic stay. If the claims diverted to the PRRA are disproportionately of the kinds in which errors are most often made and corrected, the safeguards are being taken from the claimants who need them most.

---

[100] See e.g. Karen Musalo & Marcelle Rice, "The Implementation of the One-Year Bar to Asylum" (2008) 31:2 Hastings Int'l & Comp L Rev 693 at 716-718; Lindsay M Harris, "The One-Year Bar to Asylum in the Age of the Immigration Court Backlog" (2016) 2016:6 Wis L Rev 1185 at 1196-1197; Victoria Neilson & Aaron Morris, "The Gay Bar: The Effect of the One-Year Filing Deadline on Lesbian, Gay, Bisexual, Transgender, and HIV-Positive Foreign Nationals Seeking Asylum or Withholding of Removal" (2005) 8 NY City L Rev 233 at 262-269.

[101] *INA*, supra note 94 at 8 USC § 1231(b)(3).

## 3. The Study

This Part takes up the study's two research questions and answers them from the published record of Canadian refugee adjudication. Doing so also involves answering two additional descriptive questions: how common late claims are in the published record, and how late claims fare overall at each level of the system, at first instance, on appeal to the RAD, and on judicial review in the Federal Court.

Answering these questions across tens of thousands of decisions requires reading them, and for the purposes of this study most of the reading was done using artificial intelligence. What follows describes the data, the categorization scheme, the categorization pipeline and the checks on it, the findings, and the limitations. The computer code, the categorized data, and the validation materials are published as five notebooks, referred to below as Appendices 1 through 5, and the decision texts come from a public dataset frozen at a single revision to facilitate future replication.[102]

### *3.1 The data*

The decisions used for this study come from the A2AJ's Canadian Case Law dataset, an open access collection of Canadian court and tribunal decisions.[103] Four sources were used, summarized in Table 1. The first is the body of first-instance RPD decisions that the Board has made public: 6,198 decisions from 2006 to 2020. The Board publishes only a small fraction of its first-instance reasons, the published set leans heavily negative (partly because, for much of this period, the Board did not typically issue written reasons when it granted protection), and the Board stopped publishing RPD decisions in 2020. The second is the Refugee Law Lab Reporter (RLLR),[104] which publishes positive RPD reasons obtained through access to information requests: 927 decisions from 2019 to 2024, all of them positive. The third is the RAD: 14,156 decisions published by the Board from 2013 to 2025.[105] The fourth is the Federal Court. Because the Court hears cases on every subject, its decisions were first filtered to those mentioning the RPD, the RAD, or the Immigration and Refugee Board – 11,181 decisions from 2006 to 2025 – and each of those was then screened with a threshold question, described below, asking whether the decision under review was actually an RPD or RAD refugee determination, rather than, say, a PRRA or humanitarian decision with a refugee history in the background. Through this process, 7,571 Federal Court decisions were screened in.

[102] Sean Rehaag, "Replication Appendices: Refugee Law Lab Bill C-12 Study" (2026), online: <https://github.com/Refugee-Law-Lab/rll-c-12> ["Replication Appendices"].
[103] Sean Rehaag, Simon Wallace & Clifford McCarten, "A2AJ Canadian Case Law" (2025), online: <https://huggingface.co/datasets/a2aj/canadian-case-law> (updated 2026). See also Simon Wallace & Sean Rehaag, "Introducing the A2AJ's Canadian Legal Data: An open-source alternative to CanLII for the era of computational law" (forthcoming) Canadian Journal of Law & Technology, draft available on SSRN: <https://papers.ssrn.com/sol3/papers.cfm?abstract_id=5491908>.
[104] Refugee Law Lab, "Refugee Law Lab Reporter", online: <https://refugeelab.ca/rllr>.
[105] The RAD began operating on 15 December 2012: *Protecting Canada's Immigration System Act*, SC 2012, c 17.

In all, 32,462 decisions from 2006 to 2025 were categorized, and 28,852 of them – every tribunal decision, plus the Federal Court decisions that passed the screen – form the dataset for this study.[106]

**Table 1. The corpus**

| Source | Which decisions appear | Years | Decisions categorized | In the analysis |
|---|---|---|---|---|
| First-instance RPD decisions published by the Board (RPD) | A small fraction of decisions; leans heavily negative | 2006–2020 | 6,198 | 6,198 |
| First-instance RPD decisions published in the Refugee Law Lab Reporter (RLLR) | Positive decisions only, obtained through access to information requests | 2019–2024 | 927 | 927 |
| Refugee Appeal Division decisions (RAD) | Mostly claimants contesting an earlier loss | 2013–2025 | 14,156 | 14,156 |
| Federal Court decisions mentioning the RPD, the RAD, or the Immigration and Refugee Board | Mostly claimants contesting an earlier loss, where leave was granted; only reviews of RPD or RAD refugee decisions are analyzed | 2006–2025 | 11,181 | 7,571 |
| **Total** | | | **32,462** | **28,852** |

Source: Appendices 3 and 5.

One feature of the dataset needs emphasis. It is built from the written reasons in published decisions, not from the full records of a random sample of claims. Each source publishes selectively: public RPD reasons lean negative, the RLLR is exclusively positive, and RAD and Federal Court decisions mostly involve claimants contesting an earlier loss (and in the case of the Federal Court, decisions are restricted to applications where leave was granted). Absolute rates in this corpus – the share of claims that are late, or the share that succeed – therefore say little about the refugee system as a whole. Comparisons between groups within a source, and within a level of the system, stand on firmer ground, and those comparisons are the basis for the study's findings.

### *3.2 What was categorized*

Every decision was categorized along three dimensions, and Federal Court decisions along a fourth. The full instructions given to the models are reproduced in the replication materials.[107]

*Federal Court screen.* For each Federal Court case a preliminary question was first asked about whether the decision under review was an RPD or RAD decision (including vacation, cessation, and reopening decisions). Cases involving non-immigration matters, PRRA decisions, humanitarian applications, visa refusals, detention reviews, and removals were screened out and not categorized further, even where the applicant's refugee history appeared in the background.

[106] Decisions dated outside 2006–2025 and decisions without English text in the dataset were excluded. The English text of each decision was the only input to the categorization.

[107] The verbatim prompts given to the production model are available in Appendix 2. Rehaag, Replication Appendices, supra note 102.

*Timing.* Was the claim made more than one year after the principal claimant's first entry to Canada? This question tracks the new statutory test: first entry means the first time the claimant ever entered Canada, in any status – including a visit years earlier – and leaving and returning does not reset the clock. Decisions were categorized as late where the text, or a reasonable inference from it, made clear that the claim came more than a year after first entry; as timely where they clearly came within a year; and as unclear otherwise. The instructions called for reasonable inference: a claim made at a port of entry, for example, was to be treated as timely even where the decision gives no claim date. Unclear is a common category, because published tribunal reasons redact identifying details, often including dates, and because the Federal Court often omits reciting facts that do not drive its analysis.

*Outcome.* Was the principal claimant successful? Success is always assessed from the claimant's perspective: at first instance, a grant of protection; at the RAD, an allowed appeal (or, where the Minister was the appellant, a dismissal); at the Federal Court, judicial review granted and the matter sent back (or a Minister's application dismissed). Procedural rulings that did not dispose of the claim were categorized as unclear.

*Basis.* Did the claim involve allegations of gender-based persecution, SOGIE-based persecution, both, or neither? Gender-based persecution covers domestic and intimate partner violence, forced marriage, female genital mutilation, honour-based violence, sexual violence, forced abortion or sterilization and other reproductive coercion, gender-specific dress codes and social norms, and other forms of persecution of women as a particular social group. SOGIE covers sexual orientation, gender identity, and gender expression. The categories capture what was alleged, not what the tribunal found credible, and they consider the principal claimant together with any associated claimants in the same decision. The residual categories are other grounds only (political opinion, religion, ethnicity, criminal gangs, and so on) and unclear.

### *3.3 Method: machine categorization and verification*

Large language models – the technology behind artificial intelligence tools such as ChatGPT and Claude – can now reliably extract information from legal texts, provided that the results are tested through a human review process.[108] The method used to do so for this study has three layers: a benchmark, used to test whether a smaller model could be trusted to come to conclusions similar to those of a larger model; the smaller model itself, which categorized the full corpus; and a human review of a sample of the results. What follows is an overview written for non-technical readers. The replication appendices contain the full technical detail, and everything needed to re-run the study.

[108] See e.g. Holli Sargeant, Ahmed Izzidien & Felix Steffek, "Topic classification of case law using a large language model and a new taxonomy for UK law: AI insights into summary judgment" (2026) 34 Artificial Intelligence and Law 443; Sean Rehaag, "Luck of the Draw III: Using AI to Extract Data About Decision-Making in Federal Court Stays of Removal" (2024) 49:2 Queen's Law Journal 73; Jonathan H Choi, "How to Use Large Language Models

The first layer is the benchmark, which has the narrow purpose of testing whether a smaller, cheaper model could produce the same answers as the strongest available large model. The benchmark sample is 600 decisions drawn at random from the corpus – 400 split roughly evenly among the RAD, RPD, and RLLR sources, and 200 from the Federal Court.[109] Each was categorized by Claude Fable 5, the most capable publicly available model at the time the benchmark was coded, which read every decision in full and applied the written instructions.[110] For each answer it recorded a short note identifying the passage of the decision that supports it, so that any categorization can be audited against the text. Across the 600 decisions, this process identified 46 late claims, 337 timely claims, and 172 unclear; 107 gender-based claims, 64 SOGIE claims, 11 involving both, 353 on other grounds, and 20 unclear; and 45 Federal Court decisions that were not refugee judicial reviews at all.[111]

The full corpus was then categorized by a second, smaller model – Qwen3.5, an open-weights model that anyone can download – running on a single server at the Refugee Law Lab.[112] An open model run locally has two advantages for work of this kind: categorizing tens of thousands of long documents costs almost nothing, and because the model itself is fixed and publicly available, other researchers can re-run the study against the identical model, which is not true of commercial systems that change behind their interfaces. Each decision was processed in a single call answering all the questions at once, with settings chosen to make runs as repeatable as possible. The exact instructions are reproduced in Appendix 2.

for Empirical Legal Research" (2024) 180:2 J Institutional & Theoretical Economics 214; Neel Guha, et al, "LEGALBENCH: a collaboratively built benchmark for measuring legal reasoning in large language models" (2023) NIPS '23: Proceedings of the 37th International Conference on Neural Information Processing Systems 44123, available on arXiv: <https://arxiv.org/abs/2308.11462>.

109 Replication Appendices, supra note 102, Appendix 1. The sample was drawn in two batches of 300; the smaller model's instructions were refined against the first batch, with the second held back as a check. Fifteen of the 600 decisions fall outside the 2006–2025 study window, and all of the Federal Court decisions were drawn under the original text filter described below.

110 Anthropic, "Claude Fable 5" (9 June 2026), online: <https://www.anthropic.com/news/claude-fable-5-mythos-5>.

111 Replication Appendices, supra note 102, Appendix 1.

112 Qwen Team, "Qwen3.5-122B-A10B" (24 February 2026), online: <https://huggingface.co/Qwen/Qwen3.5-122B-A10B>. The study used a 4-bit quantized version of this model (Qwen/Qwen3.5-122B-A10B-GPTQ-Int4): Replication Appendices, supra note 102, Appendix 3.

Before the full run, the smaller model categorized the same 600 benchmark decisions, and its answers were compared with the benchmark. Table 2 summarizes the results. On the Federal Court screening question, the two models agreed in all 200 cases. On outcomes, they disagreed once in 555 decisions. On the basis of the claim, they agreed 96 percent of the time – and 97 percent on the question that matters most for this study, whether any gender or SOGIE allegation is present – with the few disagreements arising mostly in mixed cases (e.g. a sexual assault in police custody inside a political-opinion claim), where the smaller model was somewhat more inclined than the benchmark to find a gender-based allegation. Timing was the hardest question, at 91 percent agreement. The direction of the timing disagreements, however, matters. When the smaller model called a claim late, the benchmark agreed about 95 percent of the time; and of the 46 claims the benchmark identified as late, the smaller model found 37, describing most of the remainder as unclear. The late-claim group is therefore undercounted rather than inflated relative to the benchmark. Most of the timing disagreements (41 of 52) were borderline calls between timely and unclear, where inferences could reasonably go either way.[113]

**Table 2. Agreement between the models on a random sample of 600 decisions**

| **Question** | **Agreement with the benchmark** |
|---|---|
| Federal Court screen: was the decision under review an RPD or RAD refugee decision? | 100% (200 of 200) |
| Timing: late, timely, or unclear? | 90.6% (503 of 555) |
| Outcome: was the claimant successful? | 99.8% (554 of 555) |
| Basis: gender, SOGIE, both, or other grounds? | 96.4% (535 of 555) |
| ***Late claims*** | |
| When the smaller model said "late", how often the benchmark agreed | 94.9% (37 of 39) |
| Of the claims the benchmark said were late, how many the smaller model also found | 80.4% (37 of 46) |
| ***Any gender or SOGIE allegation*** | |
| Overall agreement | 97.5% (541 of 555) |
| When the smaller model found such an allegation, how often the benchmark agreed | 93.8% (180 of 192) |
| Of the claims where the benchmark found such an allegation, how many the smaller model also found | 98.9% (180 of 182) |

Source: Appendix 2.

Note: Agreement is the share of decisions on which the smaller model gave the same answer as the benchmark. The Federal Court screen is scored on the Federal Court decisions in the sample; the other questions only on the decisions that were not screened out.

[113] Replication Appendices, supra note 102, Appendix 2. The characterization of the disagreements draws on a case-by-case review carried out by the benchmark model itself, which also identified at least one benchmark answer that appears wrong under the study's own instructions.

The full run followed: 30,467 decisions on July 22–23, 2026, taking about seventeen hours on a single server, and a further 1,995 Federal Court decisions on August 12, 2026, after the Federal Court text filter was widened for reasons explained below. Every one of the 32,462 decisions returned a complete, usable answer.[114] In the course of that run, 585 of the benchmark decisions were categorized a second time with identical settings, and the answers did not always match: they were the same in 94 percent of decisions on timing (most of the changes being between timely and unclear) and more often on the other questions. A re-run of the study should therefore reproduce its numbers closely rather than exactly.

The third layer is human review, conducted by the author. One hundred decisions were drawn at random from the corpus, excluding the 600 benchmark decisions, so that the review probed answers that had never been checked. The draw produced 41 Federal Court, 40 RAD, 16 RPD, and 3 RLLR decisions. Each was read in full, and every categorization was tested against the question: is this a reasonable reading of the decision? The standard was deliberately deferential. It was not "is this how I would have categorized it", because many decisions sit on borderlines – missing or redacted information, intersecting grounds – where reasonable readers could differ. A categorization failed only where it was not a defensible reading of the text.[115]

The result was 292 defensible categorizations out of 293. Every screening answer (41 of 41), every timing answer (84 of 84), and every outcome answer (84 of 84) passed. (Sixteen Federal Court decisions had been screened out, so only the screening answer was reviewed for those.) One basis answer did not: a decision categorized as "other" where a gender-based allegation clearly arose in the reasons. One further basis answer gave the author pause: a decision categorized as "gender" that involved a male claimant and forced sterilization but was not framed in gender terms by the parties or the decision-maker. It was upheld, because the instructions treat forced sterilization and other reproductive coercion as gender-based persecution. A recurring borderline involved timing: in eleven decisions, typically ones giving an entry date but no claim date, the author would have answered "unclear" but accepted the model's "timely" as a defensible inference, a tendency that, if anything, undercounts late claims.

Late claims, however, are only 8.5 percent of the corpus, so, unsurprisingly, the random sample contained just eight of them, which seemed to be thin support for the category on which the study's key findings rest. A second, targeted human review therefore examined 25 additional decisions drawn at random from those the model had categorized as late. Twenty-four of the 25 were upheld as reasonable; one was judged unclear rather than late. Putting the two human review samples together: 32 of 33 late-claim categorizations were found reasonable.

---

[114] Ibid, Appendix 3. Twenty-three decisions needed a second attempt (three of them a third, with the model's "thinking" turned off), and one exceptionally long Federal Court judgment (which the model screened out) had to be truncated. The comparison between the two runs is also set out there.

[115] Ibid, Appendix 4. Exact 95 percent confidence intervals: 98.1 to 100 percent for the 292 of 293 categorizations, and 84.2 to 99.9 percent for the 32 of 33 late-claim categorizations discussed below.

**Table 3. The human reasonableness review**

| Question | Categorizations judged reasonable | |
|---|---|---|
| | Random sample (100 decisions) | Targeted sample (25 decisions categorized as late) |
| Federal Court screen | 100% (41 of 41) | 100% (8 of 8) |
| Timing | 100% (84 of 84) | 96% (24 of 25) |
| Outcome | 100% (84 of 84) | 100% (25 of 25) |
| Basis | 98.8% (83 of 84) | 100% (25 of 25) |
| Of the timing answers: claims categorized as late | 100% (8 of 8) | 96% (24 of 25) |

Source: Appendix 4.

Note: The standard was whether each categorization was a reasonable reading of the decision, not whether the reviewer would have categorized it the same way. In the random sample, 16 of the 41 Federal Court decisions were screened out, so only the screening answer was reviewed for those. The targeted sample was drawn from decisions the model categorized as late; it is a check on the late category specifically and is not combined with the random sample, except on late-claim categorizations: across the two samples, 32 of 33 were judged reasonable.

Human review also caught a gap in the corpus. A manual check of 50 Federal Court decisions excluded through the original text filter (which screened in decisions that mentioned the RPD or the RAD) turned up one refugee judicial review that referred only to "the Immigration and Refugee Board", without naming a division. Follow-up checks suggested that roughly 5 to 7 percent of Federal Court refugee judicial reviews – nearly all from 2006 to 2016, when judgments often did not name the division – were being missed by the initial text filter. The filter was widened accordingly, the 1,995 newly captured decisions were categorized with the identical model and instructions, and the screening question filtered out the non-refugee matters that the wider net admitted. Of the 1,995, the model screened in 397 as refugee judicial reviews, concentrated in the early years (about 5 percent of the Federal Court decisions analyzed).

The human review is the most important check on the accuracy of the AI categorization. The author read 125 decisions in full and found the model's categorizations defensible in all but two instances. But it is worth keeping in mind that there was only one reviewer, who knew what the study expected to find, and the standard was reasonableness rather than correctness. The earlier benchmark was itself produced by an AI model rather than by human experts, so it shows whether the smaller model can stand in for a much stronger one, not how accurate either is against a human gold standard.

Taken together, the verification layers establish that the smaller model agrees with a far stronger model at high rates and that human review finds the smaller model's categorization to be reasonable at high rates. Readers who want more – the full instructions, the complete agreement tables, and the data and computer code needed to reproduce every number – will find it in the replication appendices.

### *3.4 Findings*

*How many claims are late*

Of the 28,852 decisions in the analysis, 2,455 (8.5 percent) were categorized as late, 18,597 (64.5 percent) timely, and 7,800 (27.0 percent) unclear. Among decisions where the timing was clear, then, 11.7 percent of claims were made more than a year after first entry, about one in nine, and 88.3 percent were made within a year. The proportion is of the same order across sources, though highest in the RLLR: 10.4 percent in the published RPD decisions, 12.8 percent at the RAD, 10.4 percent in the Federal Court, and 16.6 percent in the RLLR. It would seem, then, that the government's contention that a significant majority of claimants claim within their first year is borne out by the published record. However, the minority is not small, covering thousands of decisions. Moreover, the true share is likely higher: the decisions were written when no one had reason to record a claimant's first entry with precision, so earlier visits that would now trigger the bar may simply go unmentioned.[116]

*Who claims late*

Table 4 reports the association at the heart of the first research question. Among late claims, 41.5 percent involved gender or SOGIE allegations; among timely claims, 31.1 percent did. The odds ratio is 1.58 – that is, the odds that a late claim involves such allegations are about one and a half times the odds for a timely claim – and the probability of a gap this large arising by chance in a sample this size is negligible.[117] Read from the other direction, 15.0 percent of gender/SOGIE claims with clear timing were late, against 10.1 percent of other claims, roughly one in seven as against one in ten. Pooled across sources, the association holds for the two components separately and is stronger for SOGIE claims. A gender-based allegation was present in 29.6 percent of late claims and 23.9 percent of timely claims (odds ratio 1.34); a SOGIE allegation in 13.4 percent of late claims and 8.2 percent of timely claims (odds ratio 1.73).

[116] Replication Appendices, supra note 102, Appendix 5.
[117] An odds ratio compares the odds that a late claim involves a gender or SOGIE allegation with the same odds for a timely claim; 1.0 means no association. The p-values are from Fisher's exact test on the two-by-two table; for the pooled comparison p is on the order of $10^{-24}$.

**Table 4. Gender/SOGIE allegations by claim timing**

| Comparison | Late claims | Share of late claims with gender/SOGIE at issue | Share of timely claims with gender/SOGIE at issue | Odds ratio | Statistically significant? |
|---|---|---|---|---|---|
| ***A. Any gender or SOGIE allegation, by source*** | | | | | |
| All sources | 2,455 | 41.5% | 31.1% | 1.58 | Yes (p < 0.001) |
| RPD (published by the Board) | 531 | 41.4% | 28.1% | 1.81 | Yes (p < 0.001) |
| RLLR | 88 | 61.4% | 52.4% | 1.44 | No (p = 0.13) |
| RAD | 1,236 | 41.7% | 35.0% | 1.33 | Yes (p < 0.001) |
| Federal Court | 600 | 38.3% | 25.5% | 1.81 | Yes (p < 0.001) |
| ***B. All sources, by type of allegation*** | | | | | |
| Gender-based allegation | 2,455 | 29.6% | 23.9% | 1.34 | Yes (p < 0.001) |
| SOGIE allegation | 2,455 | 13.4% | 8.2% | 1.73 | Yes (p < 0.001) |

Source: Appendix 5.

Note: The odds ratio compares the odds that a late claim has gender/SOGIE at issue with the same odds for a timely claim; 1.0 means no association. p is the probability that a gap this large would arise by chance; "statistically significant" means that this probability is below conventional thresholds. Timely claims: 18,597 pooled (RPD 4,554; RLLR 443; RAD 8,419; Federal Court 5,181). Decisions with unclear timing (7,800; 30.8 percent gender/SOGIE) are excluded from the comparison. Panel B covers the same late and timely claims as the "All sources" row of Panel A, with gender-based and SOGIE allegations shown separately; claims involving both count in each row. Decisions whose basis was unclear are counted as not involving gender or SOGIE allegations; excluding them does not materially change the results.

Because the four sources differ in what they publish, the comparison was repeated within each. The direction is the same everywhere. The odds ratio is 1.81 in both the published RPD decisions and the Federal Court, 1.33 at the RAD, and 1.44 in the RLLR (for the latter, the 88 late claims are too few for the difference to reach conventional statistical significance). Within each source, late claims are more likely than timely claims to be about gender or sexual orientation. Taken separately, the two components are less uniform. SOGIE claims are overrepresented among late claims in all four sources (narrowly missing statistical significance at the Federal Court); gender-based claims are strongly overrepresented in the published RPD decisions and at the Federal Court, but not at the RAD, the largest source, or in the small RLLR corpus (Appendix 5, S3).

The pattern is consistent with the American record described in Part 2, where women filed late at a rate 13 percent higher than men (32.7 against 29 percent). Here, among decisions with clear timing, claims involving a gender-based allegation were late at a rate 29 percent higher than claims without one (14.0 against 10.9 percent), and claims involving a SOGIE allegation at a rate 60 percent higher (17.7 against 11.1 percent).[118]

[118] Schrag, et al, supra note 86 at 702. See also text accompanying note 99.

*Do late claims succeed, and are errors against them caught*

What remains is how late claims fare once they are decided – the second descriptive question – and, with it, the second research question: whether, among late claims, refusals of gender and SOGIE claims are overturned more often than refusals of other claims. Table 5 reports success rates, from the claimant's perspective, at each level of the system, for late and timely claims, and within late claims for gender/SOGIE claims against the rest. Success serves here as the measure of error correction. In most cases it means that a refusal was set aside, though it also includes the smaller number of cases in which a Minister's appeal or application against a grant of protection was dismissed. The absolute rates are shaped by what gets published and should not be compared across levels. The comparisons that matter are within each level – and at first instance, as explained below, even those comparisons must be treated with caution.

**Table 5. Success rates for claims with a clear outcome, by level, timing, and basis**

| Level | Timing | All claims | By basis of claim | | Gender/SOGIE claims, by type | |
|---|---|---|---|---|---|---|
| | | | Gender or SOGIE allegation | Other grounds | Gender-based allegation | SOGIE allegation |
| RPD (published by the Board) | Late (more than 1 year) | **2%** (10 of 531) | **1%** (3 of 220) | **2%** (7 of 311) | **2%** (3 of 177) | **2%** (1 of 50) |
| | Timely (within 1 year) | **5%** (210 of 4,549) | **6%** (77 of 1,278) | **4%** (133 of 3,271) | **6%** (64 of 1,071) | **7%** (17 of 235) |
| RLLR (positive decisions only) | Late (more than 1 year) | **100%** (88 of 88) | **100%** (54 of 54) | **100%** (34 of 34) | **100%** (18 of 18) | **100%** (38 of 38) |
| | Timely (within 1 year) | **100%** (442 of 443) | **100%** (231 of 232) | **100%** (211 of 211) | **99%** (116 of 117) | **100%** (132 of 132) |
| RAD appeal | Late (more than 1 year) | **23%** (282 of 1,236) | **28%** (143 of 516) | **19%** (139 of 720) | **28%** (94 of 338) | **27%** (54 of 199) |
| | Timely (within 1 year) | **28%** (2,319 of 8,418) | **31%** (925 of 2,943) | **25%** (1,394 of 5,475) | **30%** (650 of 2,179) | **36%** (321 of 889) |
| Federal Court judicial review | Late (more than 1 year) | **31%** (187 of 598) | **38%** (88 of 230) | **27%** (99 of 368) | **37%** (71 of 194) | **49%** (21 of 43) |
| | Timely (within 1 year) | **36%** (1,846 of 5,175) | **36%** (481 of 1,323) | **35%** (1,365 of 3,852) | **37%** (397 of 1,080) | **34%** (94 of 275) |

Source: Appendix 5.

Note: Each cell shows the success rate, with the number of successful claims and the number of claims with a clear outcome in brackets. Success is assessed from the principal claimant's perspective (protection granted; RAD appeal allowed, including referral back; judicial review granted, or a Minister's application dismissed). The two first-instance sources select for outcomes in opposite directions – the published RPD decisions lean heavily negative, while the RLLR publishes only positive decisions (one timely RLLR claim appears as unsuccessful because the principal claimant was refused while a family member was accepted) – so they are reported separately, and first-instance success rates are not compared in the analysis. Claims involving both a gender-based and a SOGIE allegation count in each of the last two columns. "Other grounds" includes decisions whose basis was unclear. Late gender/SOGIE claims versus other late claims: odds ratio 1.60 (p = 0.0006) at the RAD, 1.68 (p = 0.005) at the Federal Court.

At first instance, the Board granted protection in 98 late claims with a clear outcome: 10 in the published RPD decisions and 88 in the RLLR. These are claims made more than a year after entry that the Board found well-founded, and that the bar would have kept from the Board entirely. Success rates at this level are not compared, however, because the two first-instance sources select for outcomes in opposite directions: the published RPD decisions lean heavily negative (a 1 to 7 percent success rate across every group in Table 5), while every RLLR decision is positive by construction. Neither source can therefore say much about how adjudicators treated late claims relative to timely ones, or gender and SOGIE claims relative to others; within the published RPD decisions, the handful of late-claim grants (10 of 531: 3 of 220 gender/SOGIE claims and 7 of 311 others) is too small to support a comparison, and within the RLLR there is nothing to compare.

What the first-instance record establishes is that the Board granted protection in at least 98 published late claims, 57 of them involving gender or SOGIE allegations – in a context where only a tiny proportion of the many tens of thousands of positive first instance refugee decisions are reported.

At the RAD, 282 of 1,236 late-claim appeals with a clear outcome ended in the claimant's favour, 23 percent, against 28 percent for timely claims. The RAD exists to catch first-instance error, and it is catching errors against late claims at a substantial rate, and at a higher rate for gender/SOGIE late claims than for other late claims (28 against 19 percent; odds ratio 1.60, $p = 0.0006$), with gender-based and SOGIE claims behaving alike (28 and 27 percent).[119] Under Bill C-12 none of these appeals would exist.

At the Federal Court, 187 of 598 late-claim judicial reviews with a clear outcome were granted, 31 percent, against 36 percent for timely claims. The gender/SOGIE gap persists among late claims (38 against 27 percent; odds ratio 1.68, $p = 0.005$). Among timely claims, by contrast, the Court's grant rate is flat across claim types, 36 percent for gender/SOGIE claims and 35 percent for the rest. The gap appears only among late claims, and the contrast between late and timely claims is statistically significant (Appendix 5, S4).

Two patterns in the outcome data cut the other way. First, late claims succeed less often than timely claims at both the RAD (23 against 28 percent; odds ratio 0.78, $p < 0.001$) and the Federal Court (31 against 36 percent; odds ratio 0.82, $p = 0.03$). The published RPD decisions point the same way (2 against 5 percent). The deficit sits among claims without gender or SOGIE allegations, which make up nearly three in five late claims. These claims succeed less often than timely claims on the same grounds, both at the RAD (19 against 25 percent) and at the Federal Court (27 against 35 percent). Both gaps ($p < 0.001$) are as statistically robust as the gender/SOGIE differences reported above. For claims on other grounds, then, the data are consistent with the view that late claims are, on average, weaker. Late gender/SOGIE claims show no such deficit. They succeed about as often as timely gender/SOGIE claims (28 against 31 percent at the RAD; 38 against 36 percent at the Federal Court), and neither difference is statistically significant. The one exception is late SOGIE claims at the RAD, which succeed less often than timely SOGIE claims (27 against 36 percent; Appendix 5, S5). That exception is notable because SOGIE claims are the group most consistently overrepresented among late claims. The comparison with other late claims therefore shows that late gender/SOGIE claims are not weak, not that they are unusually strong. Like every outcome comparison in this Part, these figures describe appeals and judicial reviews that were brought, decided and published, not late claims generally. The argument against the bar does not, however, turn on the average late claim. The bar is categorical: it removes the hearing, the appeal and the stay from strong and weak late claims alike.

[119] Outcome is assessed from the claimant's perspective throughout, so a dismissed Minister's appeal counts as a success for the claimant. Most RAD appeals are claimant appeals.

Second, at the RAD the gender/SOGIE advantage is not specific to late claims, and the evidence that it widens with delay comes from the Federal Court. At the RAD, timely gender/SOGIE appeals also succeed more often than other timely appeals (31 against 25 percent). The gap among late claims is wider (28 against 19 percent), but the difference between the two gaps is not statistically significant ($p = 0.22$). At the Federal Court there is essentially no gap among timely claims (36 against 35 percent) and an eleven-point gap among late ones. That difference is significant ($p = 0.01$). Even there, the gap opens because late claims on other grounds fare worse than timely ones, not because late gender/SOGIE claims fare better. The pattern is consistent with first-instance decision-makers holding delay against gender/SOGIE claimants in ways that review corrects. It is also consistent with two other explanations: delay may be a weaker signal of merit in those claims than in others, or the refusals that get appealed, granted leave, or settled may differ by claim type. What the outcome data establish is that late gender/SOGIE claims fare as well on review as timely claims, and better than other late claims. They do not establish why.

*Summary*

In the published record, the answer to both of the study's research questions is yes. Claims involving gender-based and SOGIE-based persecution are overrepresented among claims made more than a year after first entry, in every source (though not to a statistically significant degree in the small RLLR corpus) and, when the sources are pooled, for each type of claim separately (although for gender-based claims alone the pattern does not appear at the RAD or in the RLLR). And among late claims, those same claims are more likely than other late claims to succeed on appeal or on judicial review, which in most cases means having a refusal set aside. In absolute terms, the published record alone contains 98 late claims accepted by the Board, 282 late-claim RAD appeals and 187 late-claim Federal Court judicial reviews decided in the claimant's favour. Late claims as a whole, however, fare somewhat worse on review than timely claims, a deficit concentrated among claims on other grounds. In considering these figures, recall that the published record is a fraction of all claims – and, in first instance decisions, skews overwhelmingly negative.

### *3.5 Limitations*

The findings should be read with five limitations in view, beyond the verification limits already discussed in section 3.3.

The first has been stated: the corpus is the published decisions, not a random sample of claims, and selection into publication differs by source. The within-source comparisons are the ones to rely on. But while within-source comparisons remove differences between sources, they do not address selection within them. Because each source contains mostly refusals or only grants, and because both delay and claim type can affect outcomes, part of the association could reflect which decisions end up in each source.

Second, the timing measure approximates whether claims are made within a year. The measure rests on whatever a decision happened to recite about when the claimant first came to Canada and when the claim was made, in reasons written before anyone had a reason to record first entry precisely. Roughly a quarter of decisions could not be resolved either way, largely because published tribunal reasons redact dates and Federal Court decisions often omit them. As sections 3.3 and 3.4 explain, both features push toward undercounting late claims. If that undercount were similar across claim types, it would tend to dilute the associations reported here rather than manufacture them. It may not be: because Guidelines 4 and 9 caution decision-makers about inferences from delay in gender and SOGIE claims, decisions in those claims may discuss, and so record, delay more often. The share of decisions with unclear timing, however, differs little between gender/SOGIE claims and other claims (Appendix 5, S7).

Third, the corpus is historical. It spans 2006 to 2025, while the bar reaches only people who entered Canada after June 24, 2020, and claimed on or after June 3, 2025. The recent intake may differ in composition from the earlier one, so the patterns documented here describe the population the system has adjudicated, not precisely the population the bar will divert.

Fourth, the categories are coarse and silent on mechanism. "Gender-based" pools domestic violence, forced marriage, female genital mutilation, and sexual violence; "SOGIE" pools sexual orientation, gender identity, and gender expression; intersecting claims are too few (1.1 percent of the dataset) to examine separately. The categorizations record what was alleged and when, not why the claim was late. Whether a given late claim reflects delayed disclosure, a change in the claimant's understanding of their own identity, a sur place development, or the expiry of another status is not something the decisions reliably say. The study establishes the pattern the advocates predicted; it does not, on its own, establish the mechanisms they propose for it, though the comparative and clinical research reviewed in Part 2 makes those mechanisms appear plausible.

Fifth, the comparisons are simple ones. Each sets two variables against each other – timing and basis, timing and outcome – with no statistical controls for country of origin, year, representation, or decision-maker; a more elaborate analysis, which the replication materials make possible, might sharpen or qualify the estimates. Nor does the study speak to what happens to late claims in the PRRA, the process to which the bar diverts them: as noted earlier, PRRA decisions are not published, so the comparison that would be most helpful – how the same claims fare on paper and at a hearing – cannot be made with the public data that exist today.

## 4. Implications and Recommendations

In twenty years of published Canadian refugee decisions, claims involving gender-based and SOGIE-based persecution are overrepresented among claims made more than a year after first entry, and among late claims they are more likely than other late claims to succeed on appeal or on judicial review. This Part discusses the implications of these findings and offers recommendations for reform.

### *4.1 The government's case, revisited*

Let us return to the three strands of the government's case and the premises underpinning them. The government contends that a significant majority of claimants make their claim within their first year, and the data agree: among published decisions with clear timing, roughly 88 percent of claims were timely. But the accompanying premise that one year is enough time for genuine refugees to come forward is contradicted by the data. About one in nine claims with clear timing in the published record was made more than a year after the date of first entry. Many of those claims – for example, 98 late claims in the dataset – were found by the Board to be well-founded. Moreover, the data demonstrate that those claims are not a random slice of the caseload. Rather, they are tilted toward claims the Board's own guidelines single out as the ones in which delay is to be expected and ought not to be held against the claimant, namely gender- and sexual orientation-based claims.

The argument about efficiency assumed that the diverted claims would have consumed Board resources without much changing outcomes. The outcome data say otherwise. The published record alone contains 282 late-claim appeals that the RAD decided in the claimant's favour, and 187 late-claim judicial reviews granted by the Federal Court – and the published record is a fraction of all decisions. Most of those decisions set aside a refusal; the remainder are cases in which a Minister's appeal or application against a grant of protection was dismissed, which the data do not separate out. The purported efficiency gains from keeping these claims away from the Board and from allowing removal prior to a judicial review outcome are real only if one is indifferent to whether the outcomes were right. The data show that the claims being diverted are disproportionately gender and SOGIE claims, the late claims on which the existing machinery of error correction has been doing the most work. Moreover, as Wallace has argued, the work will likely not disappear; it moves to PRRA officers, to stay motions, and to the Federal Court's leave docket, where 270 applications and 39 case-managed constitutional challenges had arrived within three months of Bill C-12 receiving royal assent.[120]

The argument about incentives assumed that late claims are disproportionately strategic rather than genuine, including students whose permits are expiring reaching for asylum as a fallback. The data cannot identify motives, and no doubt some claims of every kind, late and timely alike, are unfounded. But a population of predominantly strategic late claims should look weak on the merits when tested, and the published record bears this out only in part. Many late claims were granted by the Board, and while late claims on other grounds fared somewhat worse on review than timely claims, late gender and SOGIE claims did not: they succeeded about as often as timely claims, and more often than other late claims.

And the argument about adequacy is also undermined by the data, read alongside the Board's own guidelines. The reasoning in the government's Charter Statement is that the PRRA assesses the same risks the Board would have assessed, such that nothing of substance is lost.[121] That is true of the legal test and false of the process. The findings of this study suggest that late claims that will be diverted disproportionately involve allegations of gender-based and SOGIE-based persecution, claims that often turn on the credibility of the claimant's account. Those are precisely the claims that the Board has struggled to adjudicate fairly for many years, for which the Board found it necessary to write guidelines (including guidelines warning adjudicators to approach negative inferences from delay in claiming with caution), and, in this study, the late claims on which the RAD and the Court have been reversing refusals at higher rates than for other late claims. A process with no hearing by default, no appeal, and no stay is not the same process on a shorter timeline. For this population it is the removal of safeguards that the record shows are doing real work.

This study, then, offers empirical support for some of the critics' central premises: the bar will divert thousands of claims of a kind the system has found well-founded or whose refusals it has set aside; those claims are disproportionately based on gender-based violence, sexual orientation and gender identity/expression; and the bar removes the mechanisms that, in the published record, have been correcting errors in these claims.

### *4.2 Recommendations*

What, then, should be done?

[120] Wallace, supra note 57; on the litigation already filed, see "Federal Court grants request to group constitutional challenges against new asylum law", CBC News (14 May 2026), online: <https://www.cbc.ca/news/canada/bill-c-12-constitutional-challenges-9.7198216>.
[121] Charter Statement, supra note 70; see section 2.4, above (departmental counsel on *Singh*).

One possibility is to remove the one-year bar, either through further legislation or through constitutional litigation involving violations of section 7 or section 15 of the Charter.

That is the preferable outcome. It would ensure that claimants continue to have access to the regular refugee determination system. If the government is concerned that there is excessive and unwarranted resort to that system by various groups who have delayed claiming in Canada, and it has evidence to that effect, the refugee determination system already offers a solution: the government is free to intervene in these cases, present the program integrity concerns that it has about the claimants flowing from their delay in claiming, and ask that the claims be designated as manifestly unfounded – a designation that the RPD is obligated to make if the Member is of the view that the claim is clearly fraudulent. If so designated, as with the one-year bar, the claimant will not have access to the RAD or to an automatic stay pending judicial review. The difference is that the claimant will have access to an oral hearing before the RPD, and, importantly, will have the opportunity at that hearing to explain why they delayed in claiming – and will benefit from the guidelines telling adjudicators to be careful about negative inferences from delay in cases involving gender and sexual orientation.

Assuming, however, that the government is unwilling to remove the one-year bar and that the courts do not strike it down as unconstitutional, the findings of this study support other mitigating strategies.

In particular, changes should be implemented to: (a) exempt people making claims based on gender, sexual orientation, gender identity or gender expression from the one-year bar; (b) for claimants who are subject to the one-year bar, require a hearing before a claim can be denied, as Parliament has already done for claimants who are ineligible because they previously sought asylum in a country with which Canada has an information-sharing agreement; and (c), again, for claimants who are subject to the one-year bar, provide access to an administrative appeal and to a stay pending both that appeal and judicial review.

Finally, if the PRRA is to be used more frequently in Canada's refugee determination system not primarily as a final check for changed circumstances since a refugee hearing, but as many claimants' first (and only) opportunity to demonstrate that they face risks, much greater transparency is required in this form of refugee adjudication. Substantial numbers of PRRA decisions should be published. It should be easy for researchers to use access to information requests to obtain redacted versions of non-published decisions as well as granular data on outcomes. And the government should partner with outside researchers to study PRRA decision-making to ensure that it does not replicate some of the problems that have been identified in the regular refugee determination process.

## 5. Conclusion

Recall the student from Nigeria with whom this article began. He has done nothing wrong. He came to Canada with the reasonable expectation that after his studies – for which he paid much higher tuition than his Canadian peers – he would be eligible for permanent residence in Canada. That Canada changed the rules is not his fault. Moreover, it was only while he was studying in Canada that he came to identify as gay. The risks that he faces at the hands of his father – risks from which he will not be protected in Nigeria due to criminalization of same-sex sexual intimacy – arose only recently. He previously did not know that he could make a refugee claim. Under Canada's immigration legislation as revised by Bill C-12, despite having done nothing wrong, and despite having a well-founded fear of persecution, his refugee claim is ineligible for referral to the RPD, he will have a hearing only if an officer decides to hold one, he has no access to an administrative appeal, and he has no automatic stay on removal while the Federal Court decides an application for judicial review. Will he succeed with his application? It is impossible to say, but his chances are likely lower than in the regular refugee determination process.

The findings of this study say he is not an anomaly. In twenty years of published decisions, claims like his were late more often than other claims, were determined to be well-founded in dozens of published Board decisions, and, when refused, were more likely than other late claims to have the refusal set aside by the RAD or the Federal Court.

The government has said that a significant majority of claimants claim within a year. They do. The problem is that those who claim late are, like the student from Nigeria, disproportionately the people the full refugee determination system has long struggled to properly adjudicate, and, among late claimants, those whose refusals have most often been set aside on appeal and judicial review. These problems were so persistent, with errors showing up regularly in research and in judicial reviews, that the refugee determination system produced multiple iterations of guidelines aimed at improving decision-making. Those guidelines tell adjudicators not to do the very thing that the one-year bar does: infer that gender- and sexual orientation-based claims that involve delay in claiming are suspect and likely to be unfounded.

The one-year bar was a mistake, one that hopefully the government, or if it refuses, the courts, will soon correct.